\documentclass[twocolumn,prc,superscriptaddress,unsortedaddress,showpacs,preprintnumbers,amsmath,amssymb]{revtex4}

\usepackage{graphicx}
\usepackage{dcolumn}
\usepackage{bm}

\def\beq{\begin{equation}}
\def\eeq{\end{equation}}
\def\bea{\begin{eqnarray}}
\def\eea{\end{eqnarray}}

\def\fun#1#2{\lower3.6pt\vbox{\baselineskip0pt\lineskip.9pt
  \ialign{$\mathsurround=0pt#1\hfil##\hfil$\crcr#2\crcr\sim\crcr}}}

\begin{document}
\preprint{}

\title{Density dependence of nuclear symmetry energy constrained by mean-field calculations}

\author{Jianmin Dong}
\affiliation{Research Center for Nuclear Science and Technology,
Lanzhou University and Institute of Modern Physics of CAS, Lanzhou
730000, China}\affiliation{Institute of Modern Physics, Chinese
Academy of Sciences, Lanzhou 730000, China}\affiliation{Dipartimento
di Fisica and INFN-LNS, Via S. Sofia 64, I-95123 Catania,
Italy}\affiliation{China Institute of Atomic Energy, P. O. Box
275(10), Beijing 102413, China} \affiliation{School of Nuclear
Science and Technology, Lanzhou University, Lanzhou 730000, China}
\affiliation{Graduate University of Chinese Academy of Sciences,
Beijing 100049, China}
\author{Wei Zuo}\email[ ]{Corresponding Author:  zuowei@impcas.ac.cn}\affiliation{Research Center for Nuclear Science and Technology,
Lanzhou University and Institute of Modern Physics of CAS, Lanzhou
730000, China} \affiliation{Institute of Modern Physics, Chinese
Academy of Sciences, Lanzhou 730000, China}\affiliation{Dipartimento
di Fisica and INFN-LNS, Via S. Sofia 64, I-95123 Catania, Italy}
\author{Jianzhong Gu}
\affiliation{China Institute of Atomic Energy, P. O. Box 275(10),
Beijing 102413, China}
\author{Umberto Lombardo}
\affiliation{Dipartimento di Fisica and INFN-LNS, Via S. Sofia 64,
I-95123 Catania, Italy}

\date{\today}

\begin{abstract}
We establish a correlation for the symmetry energy at saturation
density $S_{0}$, slope parameter $L$ and curvature parameter
$K_{\text{sym}}$ based on widely different mean field
interactions. With the help of this correlation and available
empirical and theoretical information, the density dependent
behavior around the saturation density is determined. We compare
the results obtained with the present approach with those by other
analyses. With this obtained density dependent behavior of the
symmetry energy, the neutron skin thickness of $^{208}$Pb and some
properties of neutron stars are investigated. In addition, it is
found that the expression $S(\rho )=S_{0}(\rho /\rho _{0})^{\gamma
}$ or $S(\rho )=12.5\left( \rho /\rho _{0}\right)
^{2/3}+C_{p}\left( \rho /\rho _{0}\right) ^{\gamma }$ does not
reproduce the density dependence of the symmetry energy as
predicted by the mean-field approach around nuclear saturation
density.

\end{abstract}

\maketitle
\section{Introduction}\label{intro}\noindent
 Our knowledge about the features of nuclear matter at saturation
density $\rho_{0}$ is based primarily on the masses of nuclei,
like the density $\rho_{0}=0.16$ fm$^{-3}$, energy per particle
$a_{v}=-16$ MeV and symmetry energy $S(\rho=\rho_{0})=28-34$ MeV
\cite{PD0}. However, the variation of the symmetry energy with
density is still intensely debated \cite{DTP,DVS,PDR,JZ}. The
symmetry energy which characterizes the isospin-dependent part of
the equation of state (EOS) of asymmetric nuclear matter, plays a
crucial role in many issues of nuclear physics as well as
astrophysics. It relates the heavy ion reactions
\cite{PD,VB,BAL,JML,SKY,Feng,Yong,Ma}, stability of superheavy
nuclei \cite{JD}, fusion cross sections \cite{CR} and structures,
composition and cooling of neutron stars \cite{NS1,NS,BG,RD}. Many
theoretical and experimental efforts have been performed to
constrain the density-dependent symmetry energy
\cite{DVS,PDR,JZ,PD,VB,BAL}.

The energy per particle of nuclear matter with density $\rho =\rho
_{n}+\rho _{p}$ and asymmetry $\delta =(\rho _{n}-\rho _{p})/\rho $
is usually written as $e(\rho ,\delta )=e(\rho ,0)+S(\rho )\delta
^{2}$, where $\rho _{n}$, $\rho _{p}$ and $\rho$ are the neutron,
proton and nucleon densities. Around the nuclear matter saturation
density $\rho_{0}$, the symmetry energy $S(\rho )$ can be expanded
to second order in density as
\begin{eqnarray}
S(\rho ) &=&S_{0}+\frac{L}{3}\left( \frac{\rho -\rho _{0}}{\rho
_{0}}\right) +\frac{K_{\text{sym}}}{18}\left( \frac{\rho -\rho
_{0}}{\rho _{0}}\right)
^{2}  \label{A1} \nonumber\\
&&+\mathcal{O}\left[ \left( \frac{\rho -\rho _{0}}{\rho _{0}}\right) ^{3}%
\right] ,
\end{eqnarray}
where $L=3\rho \partial S(\rho )/\partial \rho |_{\rho _{0}}$ and
$K_{\text{sym}}=9\rho ^{2}\partial S^{2}/\partial \rho ^{2}|_{\rho
_{0}}$ are the slope and curvature parameters at $\rho_{0}$ that
govern the density dependence of $S(\rho )$ around $\rho_{0}$.
$S_{0}$ describes the symmetry energy at density $\rho_{0}$.
Recently, some progress has been made in determining the density
dependence of $S(\rho )$ around the saturation density $\rho_{0}$
from nuclear isospin diffusion, double n/p ratio in intermediate
energy heavy-ion collisions, pygmy dipole resonance, neutron skin
thickness and the nuclear binding energy. A description of isospin
diffusion data with a symmetry energy of $S(\rho )=S_{0}(\rho
/\rho _{0})^{\gamma}$ with $\gamma =0.69-1.05$ has been obtained
by using an isospin-dependent Boltzmann-Uehling-Uhlenbeck (IBUU)
transport model \cite{LWC}, and a value of $\gamma =0.5$ is
inferred from the preequilibrium neutron and proton transverse
emissions comparing to IBUU transport calculations \cite{PNT}. The
giant dipole resonance (GDR) of $^{208}$Pb analyzed with the
Skyrme interactions implies $\gamma =0.5-0.65$ \cite{GDR}. The
$S(\rho )$ extracted from more than 2000 measured nuclear masses
gives $\gamma =0.6-0.8$ \cite{ML}. More recently, the analysis of
isospin diffusion and double ratio data involving neutron and
proton spectra by an improved quantum molecular dynamics transport
model suggests $\gamma =0.4-1.05$ with $S(\rho )=12.5\left( \rho
/\rho _{0}\right) ^{2/3}+C_{p}\left( \rho /\rho _{0}\right)
^{\gamma }$ \cite{MBT}. Although significant progress was made in
determining the symmetry energy, it remains an open question
nowadays.

It has been established that the neutron skin thickness $\Delta
R_{np}$, given by the difference of neutron and proton
root-mean-square radii of heavy nuclei, correlates linearly with the
slope $L$ around the saturation \cite{BAB,ST,RJF,AWS,MC,MC2}.
Although the theoretical predictions on $S(\rho )$ with the current
nuclear mean field methods and neutron skin thickness are extremely
diverse, this correlation is universal in the realm of mean field
theory as it is based on widely different nuclear functionals
\cite{MC,MC2}. Based on the similar idea, in this work, we try to
extract a relation of the three quantities $S_{0}$, $L$ and
$K_{\text{sym}}$ in widely different mean field interactions to
constrain the density dependent symmetry energy $S(\rho)$. This work
is organized as follows. In Sec. II, we extract a relation between
$S_{0}$, $L$ and $K_{\text{sym}}$ universally within the mean field
framework since it is based on widely different nuclear mean-field
interactions. In sec. III, by employing this relation and other
considerations, we determine the density dependence of the symmetry
energy around the saturation density. With the obtained density
dependent behavior of the symmetry energy, the neutron skin
thickness of $^{208}$Pb and some properties of neutron stars are
investigated. Finally, a short summary is given in Sec. IV.

\section{Establishment of the relation for three quantities
$S_{0}$, $L$ and $K_{\text{sym}}$ }\label{intro}\noindent

Let us first establish a relation of the three quantities. With the
relation $S(\rho )=S_{0}(\rho /\rho _{0})^{\gamma }$ describing the
density dependence of the symmetry energy \cite{LWC}, one can obtain
$L=3S_{0}\gamma $ and $K_{\text{sym}}=9S_{0}\gamma (\gamma -1)$, and
thus a correlation of $S_{0}$, $L$ and $K_{\text{sym}}$ can be
derived
\begin{equation}
S_{0}=\frac{L}{3+K_{\text{sym}}/L}.  \label{A}
\end{equation}
For the other density dependent behavior $S(\rho )=12.5\left( \rho
/\rho _{0}\right) ^{2/3}+C_{p}\left( \rho /\rho _{0}\right) ^{\gamma
}$, we have $L=25+3(S_{0}-12.5)\gamma $ and
$K_{\text{sym}}=-25+9C_{p}\gamma (\gamma -1)$. Then, the correlation
of $S_{0}$, $L$ and $K_{\text{sym}}$ takes the form
\begin{equation}
S_{0}=12.5+\frac{(L-25)^{2}}{3L+K_{\text{sym}}-50}.\label{B}
\end{equation}
The shape of the density dependence of the symmetry energy $S(\rho )$ that from
the density-dependent M3Y (DDM3Y) interaction
\cite{DDM3Y} can be written as $S(\rho )=C_{k}\left( \rho /\rho
_{0}\right) ^{2/3}+C_{1}\left( \rho /\rho _{0}\right) +C_{2}\left(
\rho /\rho _{0}\right) ^{5/3}$ with
$C_{k}=(2^{2/3}-1)\frac{5}{3}\frac{\hbar ^{2}k_{F0}^{2}}{2m}=13.0$
MeV.  $S_{0}$, $L$ and $K_{\text{sym}}$ can be expressed by
parameters $C_{1}$ and $C_{2}$ with
${S}_{{0}}{=13.0+C}_{{1}}{+C}_{{2}}$, $L=26.0+3C_{1}+5C_{2}$ and
$K_{\text{sym}}=-26+10C_{2}$. Therefore, the correlation is given
by
\begin{equation}
S_{0}=2.6+\frac{L}{3}-\frac{K_{\text{sym}}}{15}.\label{E}
\end{equation}

\begin{figure}[htbp]
\begin{center}
\includegraphics[width=0.5\textwidth]{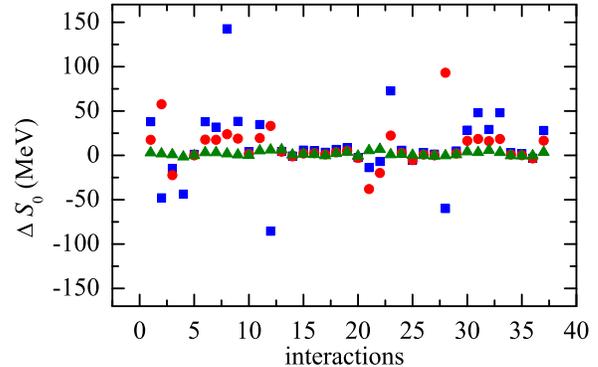}
\caption{(Color online) $\Delta S_{0}=S_{0}-L/(3+K_{\text{sym}}/L)$
(rectangle symbols), $\Delta
S_{0}=S_{0}-12.5-(L-25)^{2}/(3L+K_{\text{sym}}-50)$ (circle symbols)
and $\Delta S_{0}=S_{0}-2.6-L/3-K_{\text{sym}}/15$ (triangle
symbols) within the density dependent behavior $S(\rho )=S_{0}(\rho
/\rho _{0})^{\gamma}$, $S(\rho )=12.5\left( \rho /\rho _{0}\right)
^{2/3}+C_{p}\left( \rho /\rho _{0}\right) ^{\gamma}$ and DDM3Y
shape, respectively. The horizontal ordinate denotes the sequence
number for the interactions mentioned in text.}
\end{center}
\end{figure}

We now test whether Eqs. (\ref{A}), (\ref{B}) and (\ref{E}) work
well or not by using widely different mean field functionals
including relativistic and non-relativistic versions. As done in
Ref. \cite{XRM}, to prevent eventual biases in our study, we avoid
including more than two models of the same kind fitted by the same
group. We also avoid models yielding a charge radius of $^{208}$Pb
away from experiment data by more than $1\%$ as in Ref. \cite{XRM}
since we will study the neutron skin thickness of $^{208}$Pb. The
interactions we used here are (1)LNS1, (2)LNS5, (3)MSL0, (4)SIV,
(5)SkT4, (6)T6, (7)SkP, (8)SkM*, (9)SkX, (10)PK1, (11)D1S,
(12)SLy4, (13)FSUGold, (14)SkMP, (15)SkI5, (16)NLSH, (17)TM1,
(18)NL3, (19)NL1, (20)Sk255, (21)DDME1, (22)DDME2, (23)DDM3Y,
(24)PC-F1, (25)Ska, (26)SV, (27)QMC, (28)MSkA, (29)SkI2, (30)MSk7,
(31)HFB-17, (32)BSk8, (33)BSk17, (34)GM1, (35)GM3, (36)Sk272,
(37)v090. The root-mean-square deviations given by Eqs. (\ref{A}),
(\ref{B}) and (\ref{E}) are 38.65 MeV, 65.29 MeV and 2.94 MeV for
$S_{0}$ value, respectively. In order to give a clearer show, we
define a quantity $\Delta S_{0}$ to describe the differences
between the left hand side and the right hand side of Eqs.
(\ref{A}), (\ref{B}) and (\ref{E}), and plot them in Fig. 1. As
can be seen, $\Delta S_{0}$ with Eq. (\ref{E}) approximates zero
for these widely different interactions while $\Delta S_{0}$ with
Eqs. (\ref{A}), (\ref{B}) tends to deviate from zero considerably.
Therefore, Eqs. (\ref{A}) and (\ref{B}) can not be taken as
accurate expressions for the description of the correlation of
$S_{0}$, $L$ and $K_{\text{sym}}$ displayed by the mean-field
calculations. It is noted that the minimum value of
$K_{\text{sym}}$ is $-9S_{0}/4\approx -72 \text{MeV}$ with $S(\rho
)=S_{0}(\rho /\rho _{0})^{\gamma }$ and $-25-9C_{p}/4\approx -69$
MeV with $S(\rho )=12.5\left( \rho /\rho _{0}\right)
^{2/3}+C_{p}\left( \rho /\rho _{0}\right) ^{\gamma }$. Many
interactions, however, provide $K_{\text{sym}} < -100$ MeV (as
will be seen in Fig. 2), which leads to large discrepancies of
Eqs. (\ref{A}) and (\ref{B}) when compared to the mean-field
predictions. This fact also indicates $S(\rho )=S_{0}(\rho /\rho _{0})^{\gamma
}$ or $S(\rho )=12.5\left( \rho /\rho _{0}\right)
^{2/3}+C_{p}\left( \rho /\rho _{0}\right) ^{\gamma }$
is not suitable to describe the density dependent behavior of the
symmetry energy around $\rho_{0}$ as predicted by the mean field
approach. In particular, they cannot describe a very soft symmetry
energy due to their monotonous increase with the density. However,
Eq. (\ref{E}) is much better to be taken as a relationship for the
correlation of $S_{0}$, $L$ and $K_{\text{sym}}$. Noting that the
$S(\rho )$ with the DDM3Y shape is still approximate, Eq.
(\ref{E}) can be further improved to obtain the least deviation.
Replacing the index $5/3$ in the DDM3Y shape by a coefficient
$\gamma$, one has $S_{0}=a+L/3 + b K_{\text{sym}}$. By performing
a least-squares fit with the calculated $S_{0}$, $L$ and
$K_{\text{sym}}$ using the interactions above, the values of the
parameters are $a=3.9199$ and $b=-0.07323$ with a rms deviation of
2.12 MeV for $S_{0}$ value which is slightly different from that
within the DDM3Y shape, and hence the relation is give by
\begin{equation}
L = -11.76 + 3S_{0}+ \frac{K_{\text{sym}}}{4.55},\label{EE}
\end{equation}
with a rms deviation of 6.35 MeV for $L$ value. This formula can
be considered as a universal one within the mean-field framework
since it is based on widely different nuclear mean-field
interactions. As a consequence, the analytical and simple
expression reported in Eq. (\ref{EE}) describes in very good
approximation the high correlations displayed between $S_{0}$, $L$
and $K_{\text{sym}}$ arising from the predictions of the
representative set of employed mean-field models. The
corresponding expression for the density dependence of the
symmetry energy is given by
\begin{equation}
S(\rho )=C_{k}\left( \frac{\rho }{\rho _{0}}\right) ^{2/3}+C_{1}\left( \frac{%
\rho }{\rho _{0}}\right) +C_{2}\left( \frac{\rho }{\rho _{0}}\right)
^{1.52}, \label{A2}
\end{equation}
where $C_{k}=17.47$ is larger than that in the DDM3Y shape, which
perhaps can be regarded as the mass in the kinetic energy being
replaced by the effective mass $m^{*}$. In this step, parameters
$C_{1}$ and $C_{2}$ remain unknown.

\section{Density dependence of the symmetry energy around $\rho_{0}$ within the obtained relations }\label{intro}\noindent

\begin{figure}[htbp]
\begin{center}
\includegraphics[width=0.5\textwidth]{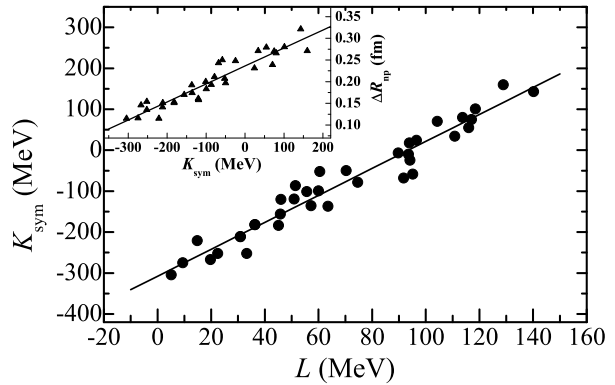}
\caption{Correlation of the curvature $K_{\text{sym}}$ with the
slope $L$ of the symmetry energy at $\rho_{0}$. The line gives the
fitting result with the correlation coefficient $r=0.972$. The
inset displays the correlation between the neutron skin thickness
$\Delta r_{np}$ in $^{208}$Pb and $K_{\text{sym}}$ value. The line
gives the fitting results with $r=0.945$.}
\end{center}
\end{figure}

\begin{table}[h]
\label{table2} \caption{The $S_{0}$ values obtained from various
independent studies in other references.}
\begin{ruledtabular}
\begin{tabular}{llllllllllllllll}
Reference  & $S_{0}$ (MeV) & Reference  & $S_{0}$ (MeV)  \\
\hline
Ref. \cite{LWC,DVS} & 31.6 & Ref. \cite{ML} & $31.1\pm1.7$ \\

Ref. \cite{MBT} & 30.1 & Ref. \cite{DL} & $32.4\pm1.1$ \\

Ref. \cite{AK}  & $32.0\pm1.8$  &Ref. \cite{HH} & 32.0 \\

Ref. \cite{TD0} & $30.048\pm0.004$ & Ref. \cite{PDR} & $32.3\pm1.3$\\

\end{tabular}
\end{ruledtabular}
\end{table}

Of the three quantities $S_{0}$, $L$ and $K_{\text{sym}}$, $S_{0}$
value is relatively well-known. Table I lists the recent $S_{0}$
value coming from recent various studies. The largest range of
$S_{0}= 31.6\pm2.2$ MeV from Table I will be used in the present
study. Yet, we have to call for an additional condition to
constrain the detailed $L$ and $K_{\text{sym}}$ values. Recently,
Centelles {\it et al.} found that the symmetry energy
(coefficient) $a_{\text{sym}}(A)$ of a finite nucleus with mass
number $A$ is approximately equal to the symmetry energy $S(\rho
_{A})$ of nuclear matter at a reference density $\rho _{A}$,
namely, $S(\rho _{A})=a_{\text{sym}}({A})$ \cite{MC}. For a given
nucleus, the $\rho _{A}$ is determined, such as $\rho _{A}=0.1$
fm$^{-3}$ for $^{208}$Pb. Here, this relationship is employed to
investigate the correlation between $K_{\text{sym}}$ and $L$.
$a_{\text{sym}}(A)$ of a finite nucleus is given by
\begin{equation}
a_{\text{sym}}(A)=\frac{S_{0}}{1+\kappa A^{-1/3}},\text{with }\kappa =\frac{9%
}{4}\frac{S_{0}}{Q}, \label{F}
\end{equation}
where $Q$ is the surface stiffness that measures the resistance of
the nucleus against separation of neutrons from protons to form a
neutron skin. In Ref. \cite{MC2}, it is shown that $S_{0}/Q$ ratio
displays a linear relationship with $L$. Then, parameter $\kappa$
can be written as $\kappa =mL+n$. By combining Eq. (\ref{F}) and
the expression of the symmetry energy obtained with the DDM3Y
shape aforesaid, the relationship of $S(\rho
_{A})=a_{\text{sym}}(A)$ can be converted into
$L^{2}+a_{1}LK_{\text{sym}}+a_{2}K_{\text{sym}}+a_{3}L+a_{4}=0$
with new coefficients $a_{1}, a_{2}, a_{3}$ and $a_{4}$ by
replacing $S_{0}$, $C_{1}$ and $C_{2}$ by $L$ and
$K_{\text{sym}}$. We fit the results from the nuclear mean field
calculations with the interactions mentioned above. However, in
the fitting process, it is found that the first two terms can be
neglected in the region under consideration. No matter whether or
not the first two terms are taken into account, one can obtain the
same root-mean-square deviation $\sqrt{<\sigma ^{2}>}=29.8$ MeV
for $K_{\text{sym}}$ value. Accordingly, this correlation can be
further simplified as a linear relation, as visibly shown in Fig.
2. By performing a two parameter fitting to this relationship, we
have
\begin{equation}
K_{\text{sym}}=-307.862 + 3.292L,\label{H}
\end{equation}
with $\sqrt{<\sigma ^{2}>}=29.8$ MeV for $K_{\text{sym}}$ value
and the correlation coefficient $r=0.972$. The results of fitting
are presented in Fig. 2 by the line. One readily sees that the
predicted $K_{\text{sym}}$ with the mean field approaches varies
largely from -300 MeV to 150 MeV. By combining Eq. (\ref{EE}) and
Eq. (\ref{H}) together with $S_{0}= 31.6\pm2.2$ MeV, in the
present work, some information on the density dependence of
$S(\rho)$ is investigated.

Fig. 3 illustrates the present estimated $L$ values compared with
those from other independent approaches. One can find that the
present finding has a remarkable overlap with but on the whole
slightly softer than the very recent results of Carbone {\it et al.}
\cite{PDR}, Liu {\it et al.} \cite{ML} and Tsang {\it et al.}
\cite{MBT}. This, to a large extent, perhaps stems from the relation
of Eq. (\ref{EE}) based on the formula of $S(\rho )$ that can
characterize the very soft symmetry energy in a much more reasonable
manner. In the other analysis with $S(\rho )=S_{0}(\rho /\rho
_{0})^{\gamma}$ or $S(\rho )=12.5\left( \rho /\rho _{0}\right)
^{2/3}+C_{p}\left( \rho /\rho _{0}\right) ^{\gamma}$, the slope
parameter of the symmetry energy at saturation density should be
overestimated. To give a clearer explanation, we present an example
here to show the overestimation of the $L$ values with $S(\rho
)=S_{0}(\rho /\rho _{0})^{\gamma}$. In Ref. \cite{DL}, Danielewicz
and Lee extracted the mass dependent symmetry energy coefficients of
finite nuclei $a_{\text{sym}}(A)=S_{0}\left( 1+\kappa
A^{-1/3}\right) ^{-1}$ with $S_{0}=27.39$ MeV and $\kappa=1.28$
\cite{ML2}. With the relation $a_{\text{sym}}(A)=S(\rho _{A})$
proposed in Ref. \cite{MC}, if the density dependent behavior
$S(\rho )=S_{0}(\rho /\rho _{0})^{\gamma}$ is adopted, one obtains
$L=34$ MeV. However, if the DDM3Y shape or the modified one (Eq.
(\ref{A2})) is applied, we obtain $L=21$ MeV and $L=24$ MeV,
respectively, lower than that stemming from $S(\rho )=S_{0}(\rho
/\rho _{0})^{\gamma}$. With $S_{0}=29.4-33.8$ MeV in Table I, the
$L$ and $K_{\text{sym}}$ values obtained with the present method are
$56\pm24$ MeV and $-125\pm79$ MeV, respectively. It is interesting
to see that the window reported in reference of Warda {\it et al.}
\cite{MC2} for the analysis of anti-protonic atoms data and that in
the present manuscript are exactly the same, as shown in Fig. 3. The
$S_{0}$ value with less uncertainty will lead to narrower windows of
$L$ and $K_{\text{sym}}$ values within our approach. The errors for
$L$ and $K_{\text{sym}}$ arising from the errors of the parameters
in the fitting are less important compared with these resulting from
the uncertainty of $S_{0}$ value. Therefore, for simplicity and
clarity, we only selected the optimal fitting results. The small
deviation and large correlation coefficient indicate that the
parameters are well constrained by the information used in the
fitting procedure.

\begin{figure}[htbp]
\begin{center}
\includegraphics[width=0.45\textwidth]{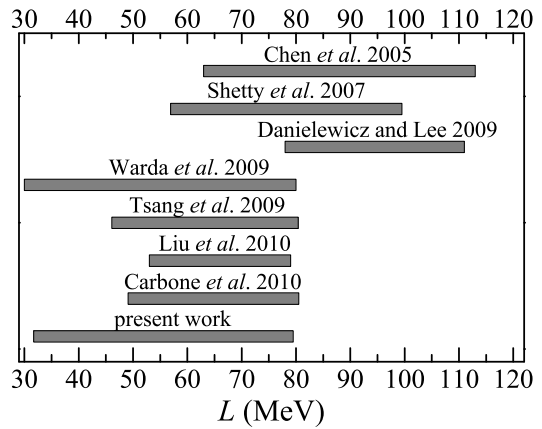}
\caption{Comparison between the $L$ values obtained in the present
work and those from other recently various analyses including
Carbone {\it et al.} \cite{PDR}, Liu {\it et al.} \cite{ML}, Tsang
{\it et al.} \cite{MBT}, Warda {\it et al.} \cite{MC2},
Danielewicz and Lee \cite{DL}, Shetty {\it et al.} \cite{DVS} and
Chen {\it et al.} \cite{LWC}.}
\end{center}
\end{figure}

It is well known that $L$ shows a linear relationship with neutron
skin thickness $\Delta R_{np}$ of finite nuclei. In Ref.
\cite{XRM}, this linear correlation is given by $\Delta
R_{np}=0.101 + 0.00147L$ for $^{208}$Pb. Because of Eq. (\ref{H}),
$K_{\text{sym}}$ should also show a linear correlation with the
neutron skin thickness $\Delta R_{np}$. Using the aforesaid
interactions for finite nuclei, we obtain the correlation between
$\Delta R_{np}$ in $^{208}$Pb and $K_{\text{sym}}$ by performing a
fitting procedure
\begin{eqnarray}
\Delta R_{np} &=&(0.236210\pm 0.003991)  \nonumber\\
&&+(0.000415\pm 0.000026)K_{\text{sym}}, \label{J}
\end{eqnarray}
where $\Delta R_{np}$ and $K_{\text{sym}}$ are in units of fm and
MeV, respectively. The result of the fitting is presented in the
inset of Fig. 2 by the line. A larger $K_{\text{sym}}$ value
implies a thicker neutron skin. The linear relation allows one to
extract $K_{\text{sym}}$ with the measured $\Delta R_{np}$.
Therefore, once an accurate measurement of $\Delta R_{np}$ is
achieved, $L$ as well as $K_{\text{sym}}$ values can be determined
simultaneously. As a consequence, a richer information about the
density dependent behavior of the symmetry energy can be achieved.
Here $K_{\text{sym}}$ is used in turn to determine the $\Delta
R_{np}$. With the $K_{\text{sym}}$ values obtained above with
$S_{0}=31.6\pm2.2$ MeV, the $\Delta R_{np}$ in $^{208}$Pb is
estimated to be $0.185\pm0.035$ fm. With the relationship between
$\Delta R_{np}$ and $L$ in Ref. \cite{XRM} together with the $L$
values of $56\pm24$ MeV that we obtained above, the $\Delta
R_{np}$ for $^{208}$Pb is $0.183\pm0.035$ fm, which is quite
consistent with that from $\Delta R_{np}-K_{\text{sym}}$
correlation. This fact indicates the justification of our approach
to a large extent. Again, our calculated $\Delta R_{np}$ for
$^{208}$Pb is in good agreement with the value from analysis of
PDR \cite{AK,PDR}. The neutron-rich skin of a heavy nucleus is
related to the properties of neutron star crusts. For instance,
the thicker the neutron skin is, the thinner the solid crust of a
neutron star \cite{NS1}. Horowitz and Piekarewicz proposed that if
$\Delta R_{np}$ for $^{208}$Pb is greater than about 0.24 fm, the
electron fraction becomes large enough to allow the direct URCA
process to cool down a $1.4M_{\odot}$ neutron star \cite{NS1}. Our
calculated $\Delta R_{np}$ is too small to allow the direct URCA
process in this canonical neutron star to occur. An almost linear
relationship between the $\Delta R_{np}$ and the critical density
$\rho_{c}$ of a phase transition from nonuniform to uniform
neutron rich matter is put forward in Ref. \cite{NS1} with an
approximate relation of $\rho_{c}\approx0.16-0.39\Delta
R_{np}(^{208}\text{Pb})$. Using this relation combining with
$\Delta R_{np}$ in $^{208}$Pb deduced above, one obtains
$\rho_{c}\approx0.09$ fm$^{-3}$ which is in accord with
$\rho_{c}\approx0.096$ fm$^{-3}$ of the microscopic EOS of
Friedman and Pandharipande \cite{CPL}. These two examples display
the applications of the correlations we employed in understanding
the physics of compact objects.

\begin{figure}[htbp]
\begin{center}
\includegraphics[width=0.45\textwidth]{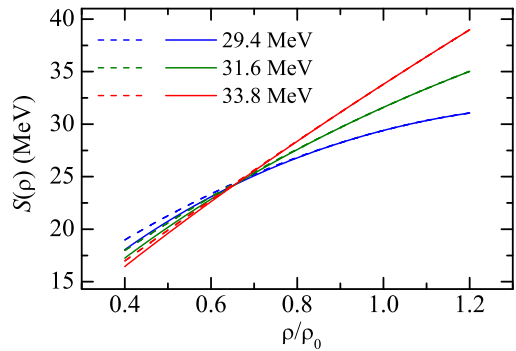}
\caption{(Color online) Behavior of the density dependent symmetry
energy with different $S_{0}$ values from Eqs. (\ref{A1}) (dash
lines) and (\ref{A2}) (solid lines), respectively.}
\end{center}
\end{figure}

When $S_{0}$ value is fixed, the parameters $C_{1}$ and $C_{2}$ in
Eq. (\ref{A2}) are accordingly determined, and hence the density
dependence of the symmetry energy in the form of a modified DDM3Y
shape is obtained. Within this density dependent behavior of
$S(\rho)$, one may investigate the symmetry energy at
subsaturation densities. Fig. 4 presents the symmetry energy
$S(\rho)$ as a function of the density with several detailed
$S_{0}$ values. The solid (dash) curves denote the calculations
with Eq. (\ref{A2}) (Eq. (\ref{A1})). As can be seen, $S(\rho)$
shows different behaviors with different $S_{0}$ values. The
symmetry energy tends to become stiffer with a larger $S_{0}$
value. When $S_{0}$ is selected to be 33.8 MeV, the relation of
$S(\rho)-\rho$ is almost linear. This explains the fact that the
relativistic mean field models have traditionally predicted a
stiff symmetry energy--these interactions give larger $S_{0}$
values. The density dependent behavior stemming from Eq.
(\ref{A2}) and Taylor expansion of Eq. (\ref{A1}) mostly coincide
with each other, which indicates that the Taylor expansion of Eq.
(\ref{A1}) can be applied in a wide density range when Eq.
(\ref{A2}) is employed for the description of $S(\rho)$. In
addition, consistent with the mean-field results of Refs.
\cite{BAB,RJF,MC}, it is found that different curves give nearly
the same value for the symmetry energy at a density around
$\rho=0.1$ fm$^{-3}$. The obtained symmetry energy $S(\rho)=
23.6\pm0.02$ MeV at a nucleon density of $\rho=0.1$ fm$^{-3}$
agrees with $23.3-24.9$ MeV by an analysis of the GDR of
$^{208}$Pb with the Skyrme interactions \cite{GDR} but is larger
than the values of $21.2 -22.5$ MeV obtained with analysis of the
GDR of $^{132}$Sn within relativistic mean field models
\cite{LGC}. Moreover, compared with that from the PDR, our result
includes a much less uncertainty. As suggested by Daoutidis and
Goriely \cite{IS}, the PDR strength measurements nowadays cannot
yield quantitative insight on the symmetry energy, and thus both
theoretical approaches as well as phenomenological interactions
need to be further improved to investigate the symmetry energy.
Some authors even proposed that the PDR strength is weakly
correlated with the neutron skin thickness of heavy nuclei
\cite{PW} and hence the density dependence of the symmetry energy.
Finally, we would note that compared with the transport models and
the PDR measurements, our approach is much more straightforward.
Yet, our approach is based only on mean field models and could only
gain some limited information about the symmetry energy. Opposite to our method, the
analysis of the experimental data within microscopic techniques, such
as the random phase approximation for the study of the PDR, may allow
one to investigate the internal structure and dynamics of the
nucleus, and hence one can obtain more detailed knowledge. These different
approaches can validate or complement each other to get more compelling results.

\section{Summary}\label{intro}\noindent
Based on the similar idea in Ref. \cite{BAB,ST,RJF,AWS,MC,MC2}
that the neutron skin thickness correlates with the slope
parameter $L$ around the saturation density in mean field models,
we have established a relation for three quantities $S_{0}$, $L$
and $K_{\text{sym}}$ in widely different mean field interactions.
With this relation and other constraint conditions, the density
dependence of the nuclear symmetry energy $S(\rho)$ has been
investigated in the present work. With the obtained density
dependence of the symmetry energy, the neutron skin thickness of
$^{208}$Pb and some properties of neutron stars were analyzed. The
main conclusions are as follows. (i) It is not suitable to take
the form of $S(\rho )=S_{0}(\rho /\rho _{0})^{\gamma }$ or $S(\rho
)=12.5\left( \rho /\rho _{0}\right) ^{2/3}+C_{p}\left( \rho /\rho
_{0}\right) ^{\gamma }$ to describe the behavior of the symmetry
energy as predicted by the mean-field approach around the nuclear
saturation density while the shape from the DDM3Y is much better.
(ii)Based on the latter formula, we have provided an analytical
and simple expression for the high correlation shown by $S_{0}$,
$L$ and $K_{\text{sym}}$ within the mean field framework.
Therefore, if the symmetry energy at saturation density $\rho
_{0}$ is known, the values of $K_{\text{sym}}$ and $L$ can be
unambiguously related by using Eq. (\ref{EE}). The $L$ and
$K_{\text{sym}}$ values in the present study are $56\pm24$ MeV and
$-125\pm79$ MeV respectively with $S_{0}= 31.6\pm2.2$ MeV yielded
in other references.(iii) The neutron skin thickness of $^{208}$Pb
displays a linear correlation with $K_{\text{sym}}$. Thus, once
the neutron skin thickness is measured accurately, not only slope
parameter $L$ but also curvature parameter $K_{\text{sym}}$ of the
symmetry energy around the saturation can be determined. Thus, a
richer information about the density dependent behavior of the
symmetry energy can be achieved. $\Delta R_{np}$ of $^{208}$Pb is
$0.185\pm0.035$ fm with the $K_{\text{sym}}$ value in item (ii)
which is too small to allow the direct URCA process in
$1.4M_{\odot}$ neutron star to occur.(iv) Within our approach, the
behavior of $S(\rho)$ against $\rho$ around the normal density
almost relies on the symmetry energy at saturation density $\rho
_{0}$. We show that a large $S_{0}$ value leads to a stiff
symmetry energy. They, however, naturally provide almost the same
result of about 23.6 MeV at $\rho =0.1$ fm$^{-3}$.

\section{ACKNOWLEDGMENTS}\label{intro}\noindent
This work was supported by the National Natural Science Foundation
of China (with Grant Nos.
11175219,10875151,10975190,11075066,11175074), the Knowledge Innovation Project (KJCX2-EW-N01) of
Chinese Academy of Sciences, the Chinese Academy of Sciences Visiting Professorship for Senior International Scientists
(Grant No.2009J2-26), CAS/SAFEA International Partnership
Program for Creative Research Teams (CXTD-J2005-1), and the Funds for
Creative Research Groups of China under Grant No. 11021504.


\begin{thebibliography}{MBT}

\bibitem{PD0}
P. Danielewicz, Nucl. Phys. {\bf A727}, 233 (2003).

\bibitem{DTP}
D. Vretenar and T. Niksi\'c and P. Ring, Phys. Rev. C {\bf
68},024310 (2003).

\bibitem{DVS}
D. V. Shetty, S. J. Yennello, and G. A. Souliotis, Phys. Rev. C {\bf
75}, 034602 (2007), and references therein.

\bibitem{PDR}
Andrea Carbone {\it et al.}, Phys. Rev. C {\bf 81}, 041301(R)
(2010).

\bibitem{JZ}
J. Zenihiro {\it et al.}, Phys. Rev. C {\bf 82}, 044611 (2010).


\bibitem{PD}
P. Danielewicz, R. Lacey, and W. G. Lynch, Science {\bf 298}, 1592
(2002).

\bibitem{VB}
V. Baran, M. Colonna, V. Greco, and M. Di Toro, Phys. Rep. {\bf
410}, 335 (2005).

\bibitem{BAL}
B. A. Li, L. W. Chen, and C. M. Ko, Phys. Rep. {\bf 464}, 113
(2008).

\bibitem{JML}
J. M. Lattimer and M. Prakash, Phys. Rep. {\bf 442}, 109 (2007).

\bibitem{SKY}
Sanjeev Kumar, Y. G. Ma, G. Q. Zhang and C. L. Zhou, Phys. Rev. C
{\bf 84}, 044620 (2011).

\bibitem{Feng}
Zhao-Qing Feng, Phys. Rev. C {\bf 83}, 067604 (2011).

\bibitem{Yong}
Gao-Chan Yong, Phys. Rev. C {\bf 84}, 014607 (2011).

\bibitem{Ma}
Chun-Wang Ma, Fang Wang, Yu-Gang Ma, Chan Jin, Phys. Rev. C {\bf
83}, 064620 (2011).

\bibitem{JD}
Jianmin Dong, Wei Zuo, and Werner Scheid, Phys. Rev. Lett. {\bf
107}, 012501 (2011).

\bibitem{CR}
C. Rizzo, V. Baran, M. Colonna, A. Corsi, and M. Di Toro, Phys. Rev.
C {\bf 83}, 014604 (2011).

\bibitem{NS1}
C. J. Horowitz and J. Piekarewicz, Phys. Rev. Lett. {\bf 86}, 5647
(2001).

\bibitem{NS}
J. M. Lattimer and M. Prakash, Phys. Rep. {\bf 333}, 121 (2000);
Science 304, 536 (2004).

\bibitem{BG}
B. G. Todd-Rutel and J. Piekarewicz, Phys. Rev. Lett. {\bf 95},
122501 (2005).

\bibitem{RD}
R. Cavagnoli, D. P. Menezes and C. Providencia, Phys. Rev. C {\bf
84}, 065810 (2011).


\bibitem{LWC}
Lie-Wen Chen, Che Ming Ko, and Bao-An Li, Phys. Rev. Lett. {\bf 94},
032701 (2005); Phys. Rev. C {\bf 72}, 064309 (2005); Bao-An Li and
Lie-Wen Chen, Phys. Rev. C {\bf 72}, 064611 (2005).

\bibitem{PNT}
M. A. Famiano {\it et al.}, Phys. Rev. Lett. {\bf 97}, 052701
(2006).

\bibitem{GDR}
L. Trippa, G. Col\`o, and E. Vigezzi, Phys. Rev. C {\bf 77},
061304(R) (2008).

\bibitem{ML}
M. Liu, N. Wang, Z. X. Li, and F. S. Zhang, Phys. Rev. C {\bf 82},
064306 (2010).

\bibitem{MBT}
M. B. Tsang {\it et al.}, Phys. Rev. Lett. {\bf 102}, 122701 (2009).

\bibitem{BAB}
B. A. Brown, Phys. Rev. Lett. {\bf 85}, 5296 (2000).

\bibitem{ST}
S. Typel and B. A. Brown, Phys. Rev. C {\bf 64}, 027302 (2001).

\bibitem{RJF}
R. J. Furnstahl, Nucl. Phys. {\bf A706}, 85 (2002).

\bibitem{AWS}
A. W. Steiner, M. Prakash, J. Lattimer, and P. J. Ellis, Phys. Rep.
{\bf 411}, 325 (2005).

\bibitem{MC}
M. Centelles, X. Roca-Maza, X. Vinas, and M. Warda, Phys. Rev. Lett.
{\bf 102}, 122502 (2009).

\bibitem{MC2}
M. Warda, X. Vinas,X. Roca-Maza, and M. Centelles, Phys. Rev. C {\bf
80}, 024316 (2009).

\bibitem{XRM}
X. Roca-Maza, M. Centelles, X. Vinas, and M. Warda, Phys. Rev. Lett.
{\bf 106}, 252501 (2011).

\bibitem{DDM3Y}
T. Mukhopadhyay, D. N. Basu, Nucl. Phys. {\bf A789} 201 (2007).

\bibitem{DL}
P. Danielewicz and J. Lee, Nucl. Phys. {\bf A818}, 36 (2009).

\bibitem{AK}
A. Klimkiewicz {\it et al.}, Phys. Rev. C {\bf 76}, 051603(R)
(2007).

\bibitem{HH}
H. Heiselberg and M. Hjorth-Jensen, Phys. Rep. {\bf 328}, 237
(2000).

\bibitem{TD0}
T. Mukhopadhyay and D. N. Basu, Acta Phys. Pol. B {\bf 38}, 3225
(2007).

\bibitem{ML2}
Ning Wang and Min Liu, Phys. Rev. C {\bf 81}, 067302 (2010).


\bibitem{CPL}
C. P. Lorenz, D. G. Ravenhall, and C. J. Pethick, Phys. Rev. Lett.
{\bf 70}, 379 (1993).

\bibitem{LGC}
L. G. Cao and Z. Y. Ma, Chin. Phys. Lett. {\bf 25}, 1625 (2008).

\bibitem{IS}
I. Daoutidis and S. Goriely, Phys. Rev. C {\bf 84}, 027301 (2011).

\bibitem{PW}
P.-G. Reinhard and W. Nazarewicz, Phys. Rev. C {\bf 81}, 051303(R)
(2010).


\end{thebibliography}
\end{document}